# dr0wned – Cyber-Physical Attack with Additive Manufacturing

Sofia Belikovetsky
*Ben-Gurion University
of the Negev*

Mark Yampolskiy
*University of South Alabama*

Jinghui Toh
*Singapore University of
Technology and Design*

Yuval Elovici
*Ben-Gurion University
of the Negev,
Singapore University of
Technology and Design*

*Abstract*—Additive manufacturing (AM), or 3D printing, is an emerging manufacturing technology that is expected to have far-reaching socioeconomic, environmental, and geopolitical implications. As use of this technology increases, it will become more common to produce functional parts, including components for safety-critical systems. AM's dependence on computerization raises the concern that the manufactured part's quality can be compromised by sabotage.

This paper demonstrates the validity of this concern, as we present the very first full chain of attack involving AM, beginning with a cyber attack aimed at compromising a benign AM component, continuing with malicious modification of a manufactured object's blueprint, leading to the sabotage of the manufactured functional part, and resulting in the physical destruction of a cyber-physical system that employs this part. The contributions of this paper are as follows. We propose a systematic approach to identify opportunities for an attack involving AM that enables an adversary to achieve his/her goals. Then we propose a methodology to assess the level of difficulty of an attack, thus enabling differentiation between possible attack chains. Finally, to demonstrate the experimental proof for the entire attack chain, we sabotage the 3D printed propeller of a quadcopter UAV, causing the quadcopter to literally fall from the sky.

## 1. Introduction

*Additive Manufacturing* (AM), often called *3D printing*, refers to the creation of 3D objects by adding thin layers, one layer at a time, to build up an object from two dimensions to three in order to create the desired form. Compared to the traditional "subtractive manufacturing" technologies, which use various cutting tools to reduce a solid block of source material to the desired shape and size, AM has numerous socioeconomic, environmental, and technical advantages. These include, but are not limited to, shorter design-to-product time, just-in-time and on-demand production, production in the proximity to assembly lines, reduction of source material waste, and especially, the ability to produce functional parts with complex internal structure and application area-optimized physical properties. These advantages have played a significant role in the increased adoption of this transformative technology in the recent years. According to the Wohlers report [1], in 2015 the AM industry accounted for $5.165 billion of revenue, with 32.5% of all AM-generated objects used as functional parts.

Due to the computerization involved in AM, several researchers have raised concerns regarding its security, including intellectual property violation [2], [3], [4], [5], [6], [7], [8] and sabotage of manufactured object's quality [9], [10], [11], [12], [13], [14]. In this paper, we focus on the latter – a threat to public safety and national security when functional parts of safety-critical systems or of critical infrastructure are sabotaged. While prior work addresses various selected aspects of this issue, to the best of our knowledge, no one has provided a holistic view and presented a proof that this type of indirect, multistage, cyber-physical attack is actually possible. This paper provides such a proof. This paper presents the complete chain of attack involving AM, beginning with a cyber attack aimed at compromising a component in the AM environment, continuing with malicious modification of a design file, leading to the manufacture of a sabotaged functional part, and resulting in the physical destruction of a cyber-physical system that employs this part. In order to accomplish this, we propose a systematic approach to identify possible attack chains that will allow an adversary to achieve his/her goals. For the selection of the attack to be performed, we propose a methodology to assess the level of difficulty of an attack. The assessment is based on the skills, tools, and network access available to an adversary.

Lastly, for the experimental evaluation of both our approach and the concern raised, we demonstrate an attack on a benign desktop 3D printer owner who prints a replacement propeller for his quadcopter UAV. Our *remote* sabotage attack causes the propeller to break during flight, and this is quite literally followed by the quadcopter falling from the sky. The remainder of the paper is structured as follows: In Section 2, we discuss the previous work in this area. Section 3 contains an overview of the Additive Manufacturing workflow which will be the basis of the rest of the study. Section 4 maps the attack chain what will lead to sabotage of a system by compromising a 3D printed functional part. We discuss the flow of the attack, tracing back from *adversary goals*, to various forms of *manipulations*, to possible *compromised elements* and *attack vectors*. In Section 5, we rank the difficulty level and impact of each attack vector by methodically deconstructing and assessing the factors involved in carrying out the attacks. Then, in Sections 6 and 7 we demonstrate an end-to-end cyber attack utilizing the methodology that we've presented and the real-life implications of this attack. Finally, we summarize our findings and discuss future work in Section 8.

## 2. Related Work

To the best of our knowledge, the very first proof of the concept that a desktop 3D printer could be compromised was presented in 2013 at the XCon2013 conference by Xiao Zi Hang (Claud Xiao). According to his keynote presentation [15], an attack can modify "printing results," including the size of the model, position of components, integrability of components, etc.

Several publications analyze the possibility of compromising 3D printers. In [12], the authors analyzed the manufacturing process chain and found several attack vectors that can be easily exploited. They focused primary on the aspects related to networks and communication. They have examined the lack of integrity checks, particularly at the stage of receiving the design (common mechanisms that are not secure include email and USB drives), the lack of physical security on machining tools, the exposure to common network attacks and the difficulty of relying on existing quality control processes. A recent publication [16] analyzes open source software that is commonly used with desktop 3D printers: *Marlin* firmware, and three GUI applications that run on PCs and communicate with the 3D printer via G-code, *Cura 3D*, *ReplicatorG*, and *Repetier-Host*. In each of these programs, static analysis of the source code and dynamic analysis of the communication between the 3D printer and the computer reveal numerous vulnerabilities that can be exploited.

Other publications dealing with this topic can be grouped based on the two main security threat categories associated with AM: intellectual property (IP) violation and sabotage of AM.

Several authors [2], [4], [8] analyze legal aspects of IP protection in AM and report numerous deficits. For instance, a 3D scan of a manufactured object is not considered an original technical drawing (blueprint) [8]; thus it can be used to legally avoid copyright protection of a blueprint.

In [6] the authors present the first published side-channel attack on a 3D printer. The authors analyze the acoustic emanations of a desktop 3D printer, and show that various sound properties can be used to distinguish between four stepper motors, three of which are used to move the printer nozzle along the *X/Y/Z* axes, while the other is used to extrude the filament while printing. The authors show that this information is sufficient to reconstruct object topology. The experimental results show an average rate of accuracy for axis prediction of 78.35% and an average length prediction error of 17.82%.

Several authors have addressed various aspects of IP protection of 3D object design. In [5], the authors survey 3D digital watermarking techniques and evaluate the potential of these techniques to withstand shape perturbations, *e.g.*, the intentional alterations or unintentional noise commonly introduced during reverse engineering. This property is necessary to trace the origin of the IP violation. In [17], the authors propose a signing methodology that aims to transition the metadata associated with the digital 3D object to the physical 3D printed object. The authors also suggest a variety of approaches to maintain provenance, including steganography, digital watermarking, content streaming and RFID hardware. Among others, the authors discuss the benefits of printing RFID tags inside the 3D objects as a way to track the objects.

However, as discussed in [3], in the context of AM, IP is not limited to the specification of the 3D object geometry; it can also include the specification of required properties (that correspond to operational parameters of a functional part), and manufacturing process parameters (which ensure that functional parts will satisfy requirements). We are not aware of any further publications addressing the latter two categories of IP in AM.

A recent article [13] generalizes the ability to sabotage AM as the weaponization of 3D printing. The authors propose a framework for the analysis of attacks involving AM and then then discuss how certain categories of attacks can generate effects comparable with those produced by weapons (*e.g.*, kinetic damage). Further, the authors argue that the targets of such an attack can be 3D manufactured objects, AM equipment, or environment.

In [11], based on an extensive survey of AM-related material science literature, the authors identified manufacturing parameters that can have a negative impact on a manufactured part's quality. The discussion focuses on AM with metals and alloys and covers a variety of AM processes, including powder bed fusion, direct energy deposition, and sheet lamination. The identified parameters include but are not limited to build direction, scanning strategy, heat source energy, etc.

For plastics 3D printers, several publications provide an experimental proof that various manipulations can reduce the part's quality. In [9], the authors developed malware to alter the STL file defining the 3D object geometry by introducing voids (*i.e.*, internal cavities) into the design. A similar approach is presented in a recent article [14]; here, the authors investigated the impact on the tensile strength of two types of manufacturing modifications: insertion of sub-millimeter scale defects in the interior of 3D printed parts, and modification of the orientation of the part during printing. As opposed to [9], in [14] defects are introduced by replacing the main material with a contaminant.

## 3. Additive Manufacturing Workflow

Figure 1 represents a high abstraction level workflow that is common in AM. This workflow represents an increasingly common scenario when AM is offered as a service[1].

---

1. As of May 2016, *3D Printing Businesses* web site (http://3dprintingbusiness.directory/) lists 851 companies offering 3D printing service.

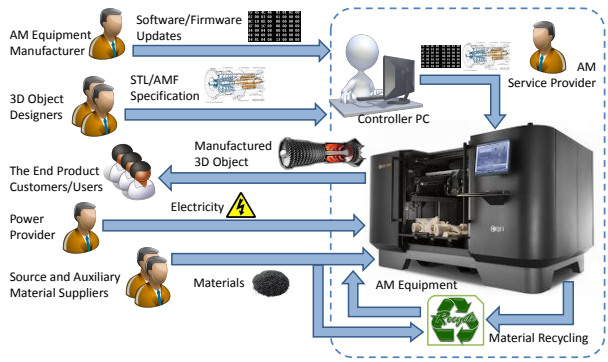

Figure 1. Additive Manufacturing Workflow

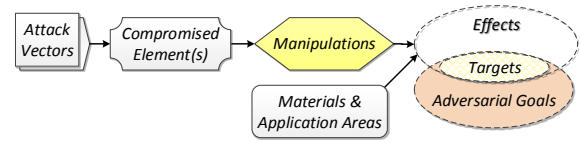

Figure 2. Attack on/with 3D printer (based on [13])

Multiple actors, the majority of whom represent enterprises, are involved in AM and provide/consume different services.

AM equipment (frequently associated with, but not limited to the actual "3D printer") is usually developed and provided by an original equipment manufacturer (OEM)[2]. The firmware and software updates (for AM equipment and the Controller PC, respectively) that extend functionality and fix bugs are provided by the OEM or companies that develop commercial software. For the desktop 3D printers, open-source software developed by the 3D manufacturers community is frequently used.

It is important to note that, for the equipment maintenance and repair (not shown in the figure), various mechanical, electrical, and electronic components (*e.g.*, motors, filters, etc.) might be required. These items are sold by OEM or third-party companies, and shipped via physical carriers.

The blueprint of a 3D object is provided either in STereoLithography (STL)[3] [18] or in Additive Manufacturing File (AMF) format [19], [20], both of which represent the Computer-Aided Design (CAD) model of the 3D object to be manufactured. The figure depicts a scenario in which object blueprints (in STL/AMF files) are provided by external 3D object designers directly to the AM service provider. Another common scenario is when the design is provided by the end product customer, who either designed the proposed object (common for enterprise customers) or bought the design/blueprint from a designer (increasingly common for individual consumers).

At the AM service provider site, a STL/AMF file can be either be directly transferred to a 3D printer (*e.g.*, via computer network or USB stick) or interpreted by the controller PC. In the latter case, the controller PC sends the 3D printer either individual control commands (in some cases encoded in "G-code" [21], a language commonly used in Computer-Aided Manufacturing, CAM[4]), or as a tool path file containing a sequence of (3D printer-specific, often proprietary) commands to be executed [9]. The transmission commonly utilizes a computer network[5].

For the actual manufacturing, AM equipment requires electricity, and a variety of source and eventually also auxiliary materials. While source materials are included in the end-product, auxiliary materials have different functions, supporting or enabling production. For instance, support structure enables the printing of complex geometries; If lasers are used as a heat source, inert gas (usually, argon) is often used, etc. Currently, the source materials for plastic printers are commonly supplied by the OEMs; the source material market for metal printers is less restrictive [1].

Depending on the AM process, the source material, and the part geometry, the production workflow can include several post-processing steps (not shown in the figure). These typically include removal of support structures used in the production of 3D objects with complex geometry. For metal parts, hot isostatic pressing (HIP), finish machining, and surface finishing are common post-processing steps. In the case of functional parts, non-destructive testing (*e.g.*, ultrasonic C-scan) is usually employed as the final step. Only after all necessary production and post-production steps are accomplished, is the manufactured 3D object is delivered to the customer via a physical carrier.

In order to reduce environmental impact and manufacturing costs, the remaining source material can be partially recycled. This is especially the case in power bed fusion[6] (PBF), an AM process that can be used for both metals and plastics. The reused powder is often sieved and mixed with "virgin" powder in proportions that minimize the negative impact on the part's quality to an acceptable level[7].

## 4. Attack Chain

Figure 2 outlines how attacks on or with AM can be performed. A variety of attack vectors can be used to compromise one or more elements of the AM workflow. The compromised element(s), their roles in the workflow, and the degree to which an adversary can control these element(s) determine the kind of manipulations an adversary can perform. In conjunction with the type of AM equipment, source materials, and the application area of the manufactured part, these manipulations determine the potential effects. Only a subset of the resulting effects may intersect with the adversarial goals. In what follows, we refer to this intersection as *attack targets* or *threats*.

---

2. In 2015, 62 system manufacturers in 20 countries produced and sold industrial-grade AM equipment and it is estimated that hundreds of small companies offer desktop 3D printers [1].

3. The abbreviation STL is overloaded with several backronyms, including but not limited to Standard Tesselation Language, Surface Tesselation Language, etc.

4. CAM is frequently referred to as *Cybermanufacturing*.

5. In the case of desktop printers, USB connection is common.

6. In the PBF processes, a thin layer of the source material (usually, metal or polymer) in powder form is distributed in a powder bed. That layer is fused by a heat source (either laser or electron beam) that melts the profile of the next slice of the 3D object. The powder distribution and fusion sequence is repeated layer by layer.

7. Due to the exposure of the unused powder to high temperatures, the properties of particles can change (in the case of plastic) and/or powder particles can agglomerate into large clusters. Both can have a negative effect on the final product quality.

In this section we use the framework depicted in Figure 2 for the identification of possible attacks involving AM. Inspired by attack trees [22], we begin with the achievable adversarial goals (or attack targets), followed by a discussion of the manipulations that are needed to achieve these goals, and then we discuss which elements in the AM workflow can exercise the manipulations listed, and finally address attack vectors that enable these elements to be compromised. The outcome of the analysis presented in this section is summarized in Figure 3.

### 4.1. Adversarial Goals

From the AM security threats identified in the research literature (see Section 2), in this paper we consider only an intentional sabotage of a 3D-printed functional part in this study.

**Sabotage of Manufactured Part:** Functional parts are typically designed to maintain specified operational conditions for an extended period of time. We can distinguish between two cases of sabotage. In the first, the part can be altered in such a way that the normal operational range exceeds its (altered) strength, thus causing the part to break under supposedly normal operational conditions. Second, the part can be altered in a way that the material fatigue develops faster than supposed, thus causing part to break sooner than expected under normal operational conditions.

### 4.2. Manipulations

The stated goal of a manufactured part's sabotage can be achieved via various manipulations. In this paper, we only consider manipulations that can be executed in the cyber domain. Manipulations are represented by Influences and Influenced elements. Influenced elements describe the object that is manipulated by an attack and Influences describe the modification that is done of the Influenced element.

Until now, the research literature has identified the following two major categories of manipulations that can influence an AM-generated part's quality: modification of the object's specification and manipulation of the manufacturing process. We discuss the Influences and the Influenced elements that are involved in those manipulations.

We further restrict our considerations to the operational phase of the manufacturing life cycle.

#### 4.2.1. Influences.

**Object Specification Modification:** Object's specification describes the object's geometry, orientation, and ultimately its material; the latter is only relevant for multi-material AM equipment. It should be noted that the object specification can have various representations, based on the "location" in the AM workflow (see Figure 1). It is commonly associated with the STL or AMF files, both of which are CAD formats, but it can also be represented within a toolpath file or as a series of individual G-code commands, etc.

- **Defects:** It is obvious that the object's geometry and material impact its mechanical properties. Changing exterior shape can affect a part's integrability in a system and eventually detected by visual inspection. Several researchers have proposed the use of internal defects as a means of sabotage [9], [11], [14]. The negative impact of internal voids (*i.e.,* cavities) and a contaminant material defects on mechanical properties have been demonstrated experimentally in [9] and [14], respectively. Further, as noted in [11], this kind of attack will eventually affect the part's weight and weight distribution – properties that can impact the performance of the system employing such a part.
- **Orientation:** In the material science, it is very well known that anisotropy[8] of 3D-printed parts is fundamental in several AM processes [23], [24]. Based on this property, researches have proposed changing the built direction as a means of sabotaging a manufactured part's mechanical properties [11]. An experimental proof of this attack was recently shown in [14].

**Manufacturing Process Manipulation:** As opposed to the traditional subtractive manufacturing, AM not only defines the manufactured object's geometry but also "creates" its material. Various parameters of an AM manufacturing process[9] influence the microstructure of the created material, thus defining its physical properties. This aspect has been intensively investigated in the material sciences from the quality assurance perspective [25], [26]. For metals and alloys, a qualitative analysis of manufacturing parameters manipulations that can be used to sabotage a part's quality was presented in [11]. Parameters like layer thickness, scanning strategy, heat source energy, etc. have been identified as potential subjects of malicious manipulations. In the case of fused deposition modeling (FDM), an AM technology that is popular with desktop 3D printers (and that we will use in the case study in Sections 6 and 7), parameters like nozzle temperature, print bed temperature, filament extrusion speed, distance between extruder and the printed object, etc. can be manipulated. All these can eventually have an impact on the strength of bonding between layers, thus impacting the part's mechanical properties. The effects of some of these manufacturing parameters manipulations have been shown in [27].

---

8. Anisotropy means that properties vary in different directions; in the discussed case, mechanical properties like tensile strength are meant.
9. These parameters and the impact of their manipulation vary greatly across different AM technologies.

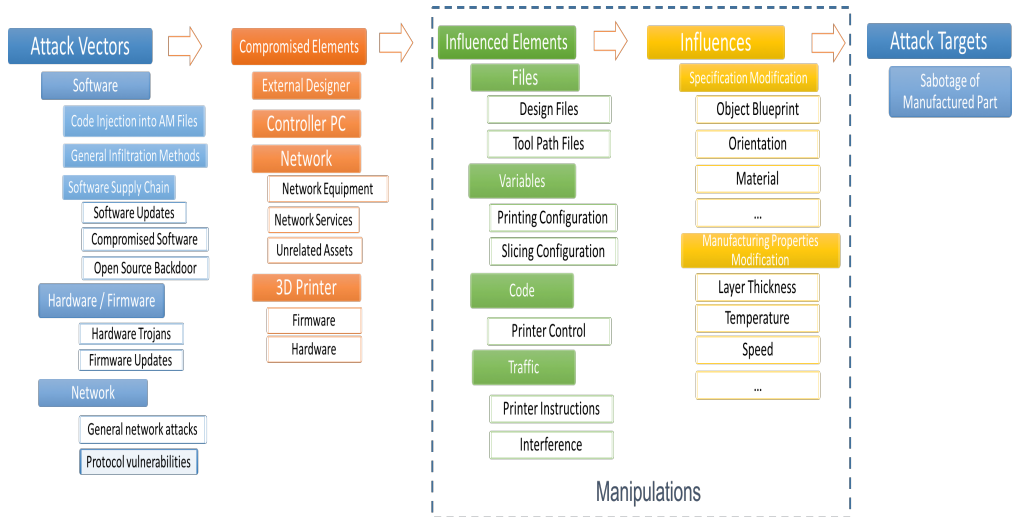

Figure 3. Attack Flow

#### 4.2.2. Influenced Elements.

**AM Files:** AM files are the input to the manufacturing process. STL/AMF files contain object's specification and any modification to these files will directly influence the resulting object. Printer instructions, that are generated from these design files, are compiled into a toolpath file that can be changed on the Controller PC, during transit or even on the 3D printer. The printer's instructions in the toolpath (commonly G-code commands) represent object's specification as well as manufacturing properties that are related to the printing process.

**Configuration:** Minor changes in configurations can impact the resulting object. Crucial properties of object's specification can be manipulated by minor changes in configuration settings in the slicing software. Properties like layer thickness, fill density, shell thickness, etc. Additionally, modification of printer configuration, like print bed temperature, nozzle travel speed, etc, can impact the manufacturing process.

**Code:** Controlling the software enables a fine-grain control over the outcome. By influencing the software that runs on the Controller PC or on the 3D printer, the adversary can achieve both manipulation categories.

**Traffic:** Controlling the traffic can cause interference with the times or content of the sent data. Some printers receive the printing instructions over the network and any disturbance or modification to this data impacts the resulting object. In [26], the authors show that even slight printing delays between each printed layer can affect the compressive strength, toughness and tangent modulus of samples printed.

#### 4.2.3. Characteristics of Manipulations.
Orthogonal to the presented categories of manipulations are characteristics of these manipulations. These depend on the subject of manipulation and also on the compromised element that exercises the change (we will discuss the latter in Section 4.3). We propose the following categorization:

**Indiscriminate/Selective:** Manipulations such as change of the object's orientation are rather indiscriminate and affect the entire object. Similarly, configuration variables that determine the slicing tool's properties will lead to rather indiscriminate changes that affect the processing of the entire design file. Other manipulations can be selective, depending on both the subject of manipulation and the compromised element that exercises this manipulation. For example, layer thickness can either be modified in the slicing software configuration and indiscriminately impact all the layers of the printed object or modified by changing a subset of G-code commands, impacting only selected layers during the manufacturing process.

**Static/Dynamic:** Another characteristics is whether manipulations are statically present or dynamically introduced. For instance, modifications of STL/AMF files are generally static in nature, and they will be replicated every time the file is used to manufacture the described object. Compromise of software or firmware involved into AM processes will enable dynamic manipulations. Modifications to the toolpath or G-code commands[10] can then be performed dynamically "on the fly" and triggered by a combination of various events. As demonstrated by the Stuxnet attack [28] has illustrated, dynamic attacks can be significantly harder to identify as an attack; therefore, such attacks can continue being active for a longer period of time undetected.

**Direct/Indirect:** In object specification manipulations (*e.g.,* in the STL/AMF file), the configuration values of the AM tools' and 3D printer's, instructions sent to the 3D printer (*e.g.,* as individual G-code commands), and status information sent back directly affect the manufactured object (and its quality) directly. In addition, as discussed in [26], [29], also the timing of particular commands and status information, as well as the power supply of the 3D printer, can have a significant impact on the manufactured process, and therefore on the manufacturing part's quality. Various classical network attacks (DoS, etc). can cause indirect manipulations, *e.g.,* by causing G-code commands to arrive too late or out of order. It should be noted that while the impact of direct manipulations on a part's quality can be predicted with a high level of certainty (and/or tested in a laboratory environment), the impact of indirect manipulations is rather stochastic.

---

10. Both toolpath and G-code commends can contain/specify information about the manufactured object as well as instructions for the 3D printer configuration; the latter will affect manufacturing process.

### 4.3. Compromised Elements

The above mentioned manipulations are only possible if at least one of the following elements of the AM workflow is compromised. Note that not all compromised elements can exercise all manipulations, and that manipulations available to different compromised elements might be comprised of different characteristics.

**External Designer:** In the considered AM workflow, the STL/AMF blueprint files are provided to the AM Service Provider either by external designers or downloaded from the Internet. In both cases, the files are received via an external (Internet) connection and serve as the basis for the entire manufacturing process. The designer is likely located outside the trusted environment; and can therefore be compromised either by a cyber attack or impersonation of a malicious actor.
In this case, the STL/AMF files originating outside the trusted environment may contain an altered object design, and both the introduction of defects and change of the orientation are possible. This represents a direct, static manipulation; whether it can be selective or not depends on the file format used.

**Controller PC:** The Controller PC operates on the STL/AMF design files, converts the design to G-code command and issues corresponding those commands to the 3D printer. If the Controller PC is not dedicated solely to the AM process and is also used for unrelated activities such as web browsing, etc. (as is commonly the case of private PCs connected to a desktop 3D printers, and cannot be completely excluded in an industrial environment either), it can be compromised. Exploitation of the Controller PC can target software that is either directly related to the AM process, *e.g.,* tools that are supporting the AM process, or unrelated software components that are executed on the Controller PC. If malicious code is running on the PC, it has direct access to, and therefore can manipulate, the design files, the generated tool path files, individual G-code commands, the 3D printer configuration, and ultimately also firmware updates for 3D printer.

**Network:** In the AM workflow, network communication is present for transferring files from an external designer of a 3D object to the Controller PC (external network communication) and also between the Controller PC and the 3D printer (internal network communication). Both can be compromised, enabling changes of the transmitted data.
We distinguish between the following network elements that can be compromised:

**Network Equipment:** Both the hardware and software of network equipment such as routers, switches, hubs, etc. can be compromised.

**Network Services:** Network communication is supported by various services including DNS, Active Directory, Mail servers, etc. Compromise of such elements can control the flow and the access on the network.

**Unrelated assets:** Other computers, network-enabled devices such as printers, and mobile devices that have access to the network can be compromised, thus enabling malicious interference into other components involved in the 3D printing process.

Depending on whether the external or internal network connection is compromised, a variety of manipulations can occur, however they cannot exceed the scope of manipulations available to the external 3D Object Designer or Controller PC, respectively. These manipulations can be further restricted, depending on which the network element that is compromised.

**3D Printer:** The 3D printer is the heart of the manufacturing process. Control over the printer can compromise the integrity of the printed object's design and manufacturing process. The attacker needs to control either the 3D printer's firmware or the hardware to achieve his/her goals. If the 3D Printer is compromised, the whole spectrum of manipulations is possible without any restrictions[11].

### 4.4. Attack Vectors

An attack vector is a path or means by which an attacker can compromise and thus gain control over an element in the AM workflow. In this paper, we only consider cyber attack vectors that can be used to compromise one or more of the elements described above. We distinguish between the following attack vectors:

**4.4.1. Software Attacks.** The Software that is used on the Controller PC or the 3D printer can be compromised and execute additional unintended functions.

It has been well demonstrated that programs are full of vulnerabilities and are vulnerable to arbitrary code execution. According to [30], the defect per KLOC stands at 6.1 for examined mission-critical software systems that were examined. Also, the Time-to-Fix can take days to months.

---
11. Here, logical restrictions are meant; all manipulations are within the boundaries of physically possible.

3D printing softwares are no exception. Since there is a variety of software that is directly or indirectly involved in the 3D printing process, there is a wide range of potential software vulnerabilities that can be exploited.

**Code Injection into AM Files:** We consider the AM design files the most vulnerable components, because they generally originate outside the controlled environment and are generated by tools that are frequently provided by third parties. The attacker can change the original design files at several steps of the AM process, starting with the external designer's site and the file's delivery to the Controller PC, up to its representation in the 3D printer itself.

If an attacker can find a vulnerability in the software involved in the manufacturing process, a specially crafted file can be provided as a source for this software, thus eventually executing an arbitrary code. There are two types of AM files: the design files and the toolpath files. Compromising of the design files (commonly of the STL/AMF type), can be done either from the external designer side, the Controller PC itself (by a malicious software already running on the PC or other compromised network components), or during the files' network transmission. A well-known example of a malicious code embedded in a CAD file is the MEDRE.A worm [31]. Eventually it is also possible to compromise the 3D printer itself, *e.g.,* by a specially crafted toolpath file.

**General Infiltration Methods:** Software attacks aim to compromise software running on one of the compromised elements. These attacks frequently exploit software vulnerabilities, *i.e.,* weaknesses in the system that commonly originate from software bugs, specifically errors, mistakes, or oversights that can result in unexpected and typically undesired behavior. Compromise of a single device in the AM network can lead to compromise of other crucial components via lateral movement. An adversary will search for the easiest and least protected point of entry by leveraging common infiltration methods such as: spear phishing via emails or fraudulent websites, malicious attachments, external devices, brute force hacking, stolen credential, etc.

**Software Supply Chain:** Software vulnerabilities can be accidentally or intentionally inserted into the software at any point in its development, distribution or use process. Software end-users have limited ways of finding and correcting these defects to avoid exploitation.

- **Compromised Software:** Software that is used on the Controller PC or 3D printer can be compromised and execute additional unintended functions. Such functions can compromise either AM-related files or the communication between AM components. The attacker can replace the software with its own malicious version or compromise specific libraries.
- **Software Updates:** Software updates are used to fix bugs that cause operational/security problems or to add new features/functionality into the program. Software updates are a potential threat, not only because of the addition of new code which might contain bugs, but also because the addition of new code introduces another attack vector to the environment. Software updates can be hijacked and manipulated if not implemented correctly. One famous example is the Flame malware that spoofed the request for Windows Updates on the local network and delivered malicious code instead [32].
- **Open Source Backdoor:** An attacker can gain access to the network by integrating a backdoor into open source software. This is not a targeted attack since it is hard to influence the open-source software that is used. Moreover, the backdoor must escape the checks and the reviews of the code's owners of the code. Nevertheless, since in 3D printing, there is a variety of open-source malwares used, this attack vector has a high likelihood of occurring.

**4.4.2. Hardware/Firmware Attacks.** The focal point of the AM process is the 3D printer. By controlling the printer, the adversary can impact the outcome of the manufacturing. Thus the adversary needs to gain control over the firmware/hardware of the 3D printer in order to fully control the printing process. Most of the printers are not directly connected to the Internet, thus the adversary needs to gain a foothold inside the network and leverage it for a secondary attack of the 3D equipment.

**Firmware/Hardware Vulnerabilities:** Similarly to software, the existence of bugs or oversights in the firmware/hardware components is inevitable. They are less common than software vulnerabilities since the functionality of the 3D equipment is limited and finding/testing the vulnerabilities requires the hardware itself or an emulator (which are still not common).

**Hardware trojans:** Hardware attacks, in the form of malicious modifications of electronic hardware at different stages of its life cycle, pose major security concerns in the electronics industry. An adversary can mount such an attack by introducing a hardware Trojan into the system, that can manipulate the AM process-related data it has access to.

**Firmware Updates:** A malicious firmware update is one of the ways in which an adversary can compromise a 3D printer or network elements.

**4.4.3. Network Attacks.** Since the AM equipment can reside on the same network as the Controller PC, traditional network attacks can impact the 3D printing process. Although most printers use propriety protocols, they are vulnerable to the same attacks as any other network devices.

**General Network attacks:** Some of the network attacks lead to full control over the communication. The attacker can secretly relay and possibly alter the communication between two parties. Attacks in which communication is fully controlled are called "man-in-the-middle" attacks, and this type of attack can include sniffing, spoofing and session hijacking. Moreover, an attacker can cause damages and delays in the manufacturing process by triggering a denial-of-service attack in the network. Such attacks can halt production and suspend the legitimate communications on the network. Any interference in the timing of the printing process can have significant effects on the quality of the resulting object [29].

**Protocol vulnerabilities:** There may be logic bugs in the protocols themselves. An attacker can hijack a communication by using logical mistakes in the protocol, as well as implementation mistakes in the software. These kinds of bugs can be extremely harmful, since changing the protocol can be expensive and time-consuming. One well known oversight is the Transaction ID Guessing attack on the DNS protocol [33].

## 5. Attack Difficulty Assessment

While in the previous section we have discussed elements that can lead to a stated goal (sabotage of a manufactured part's quality), the individual attack paths have different levels of difficulty; they may vary in the hacking skills required, AM proficiency, and required access. Therefore, in this section we assume the adversarial perspective of an attacker who wants to select a less challenging path to achieve the stated goal.

In this section we propose a methodology that takes into account the exploitation difficulty of the attack vector and the AM proficiency needed, mechanical and/or material engineering knowledge to determine what manipulation to induce. We analyze both aspect and correlate them to create a holistic approach.

For each category, we examine several leading factors and rank them based on the suggested scale. We then calculate the final score of each attack vector and manipulation based on the following equation: $\lceil \sum Scale * Weight \rceil$ where the weights of each factor is the same.

### 5.1. Attack Vector Exploitation

We examine the technological feasibility of each if the attack vectors. Since there are multiple factors that can be taken into consideration, we focus on the most significant elements to measure exploitation difficulty: the hacking skills that are needed, the availability of tools and the level of access to the AM environment. Figure 4 presents the detailed categorization of each attack vector according to the exploitation difficulty. The scales are determined in the following way:

**Hacking Skills Level:** The skill level needed determines which attack vectors are more feasible to exploit by a particular adversary. On a global scale with multiple adversaries, it also provides an indication of which attacks are more likely to occur. If the attack requires a higher level of hacking skills, it is less likely to occur. The Scale is ranging from 1-3:

| | | | Technological Difficulty | | | |
|---|---|---|---|---|---|---|
| Attack Vectors | | | Hacking Skill Level Needed (1 – Easy, 2 – Moderate, 3 – Hard) | Access to the AM Network (1 – Not Needed, 2 – Needed, 3 – Specific Needed) | Availability of tools (1 – Highly available, 2 – SW dependent, 3 – HW dependent) | Calculated Score |
| Software | | Code Injection | 2 | 1 | 2 | 1.7 |
| | | General Infiltration Methods | 2 | 1 | 2 | 1.7 |
| | Software Supply Chain | Compromised Software | 2 | 3 | 2 | 2.3 |
| | | Software Updates | 2 | 3 | 2 | 2.3 |
| | | Open Source Backdoor | 1 | 1 | 1 | 1 |
| Hardware/ Firmware | | Hardware Trojans | 3 | 3 | 3 | 3 |
| | | Firmware Updates | 2 | 3 | 3 | 2.65 |
| Network | | General Network Attacks | 1 | 2 | 1 | 1.3 |
| | | Protocol Vulnerabilities | 2 | 3 | 2 | 2.3 |

Figure 4. Technological difficulty calculation

1) Low skill level - Attacks that can be executed using widely available tools and do not require any reverse engineering skills.
2) Moderate skill level - We consider as moderate all attacks that require basic reverse engineering skills. Most of the attack vectors fall into this category.
3) Advanced skill level - We consider as advanced all attacks that require advanced hacking and reverse engineering skills both in software and hardware.

**Level of Access to the AM Environment:** This factor determines the level of access the attacker needs to gain into the AM environment in order to carry out each exploitation. Some attacks can be performed outside the AM network, either from the Internet or through the external designer's site, while others require a foothold inside the network. The Scale is ranging from 1-3:

1) Attacks that can be carried out from outside of the AM network.
2) Attacks that need some foothold in the AM network.
3) Attacks that require a foothold at a specific element in the AM network.

**Availability of Tools:** All attacks require specific tools used for testing and carrying out the attack. The simpler and more common the tools are, the easier it will be for the attacker to prepare and execute the attack. There are two types of tools needed: first, general tools that are used for hacking and second, target platform oriented tools that are needed for testing. We are focusing on the latter. The Scale is ranging from 1-3:

1) Attack vectors that require general tools or specific AM tools that are widely available on the Internet.
2) Attack vectors that require a specific version of software.

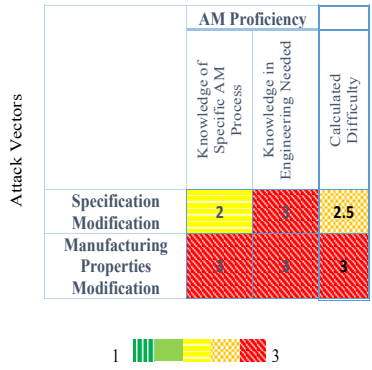

Figure 5. AM Mastery Categorization

3) Attack vectors that require a specific version of firmware and hardware.

## 5.2. Definition of Malicious Manipulation

We examine the level of AM proficiency needed in order to define the manipulation that will lead to the final goal. We calculate this based of the following factors: proficiency in the AM process and tools and the Mechanical engineering and/or Material science knowledge needed to define a manipulation that is likely to cause the desired effect. Figure 5 contains a detailed categorization of the manipulations according to the level of AM proficiency needed.

The considered factors are:

**Proficiency in the AM Process and Tools:** In order to achieve the desired goal, the adversary has to have some knowledge about the AM process, the tools that are used, and potentially the specific setup in the victim's environment. The Scale is ranging from 1-3:

1) Required basic or no knowledge.
2) Requires moderate knowledge.
3) Required advance knowledge.

**Mechanical Engineering/Material Science Knowledge:**
In order to determine the application area of the manipulation, a certain amount of Mechanical engineering and/or Material science knowledge is required. The achievable damage can vary per printed item, destined environment, and the AM equipment. The Scale is ranging from 1-3:

1) Required basic or no knowledge.
2) Requires moderate knowledge.
3) Required advance knowledge.

## 5.3. Attack Heat Map

We correlate the attack vectors and the manipulations that we've analyzed and present in Figure 6 the *heat map*

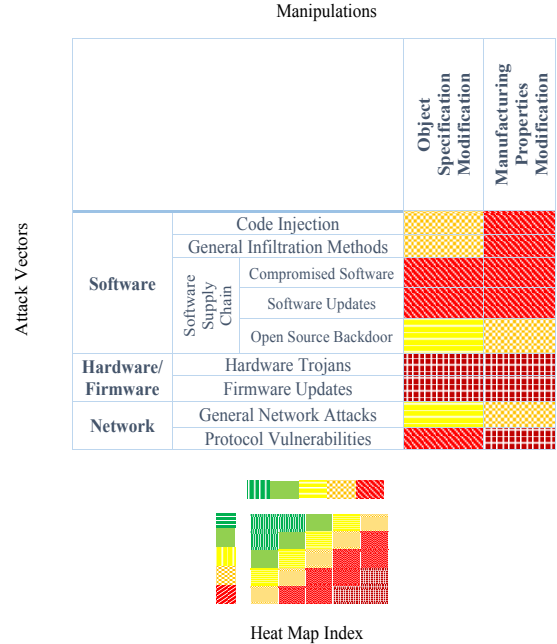

Figure 6. Attack Vector and Manipulations Heat Map

of the attacks that aim to damage functioning parts and the correlating *heat map index*. The table in the figure combines all of the previous information into one visual display. For each cell in the table, we correlate the assessment of the technological difficulty of the attack vector and the AM proficiency needed to execute the manipulation, which are calculated in Figure 4 and Figure 5 respectively. The color is calculated according to the heat map.

## 6. Case Study, Part I: Attack Preparation

This Section provides an empirical evidence that it is possible to cause physical damage of a CPS via sabotage of its 3D-printed components. To the best of our knowledge, this is the very first study that shows the entire chain of attack, beginning with a cyber attack to infiltrate a computer, and leading – over multiple stages – to the physical destruction of the victim CPS. A video demonstrating the final stage of attack can be found under [34].

In this section, we first outline the scenario considered in the case study, provide an analysis of attack options available to an adversary, and lastly, we present the attack.

### 6.1. The Scenario

**6.1.1. Victim and Adversary.** We consider a realistic scenario involving an owner (home user) of a desktop 3D printer (see Figure 7) who uses this printer to produce replacement propellers for his quadcopter UAV – a drone that he flies frequently, sometimes at a high altitude. In order to produce the propellers the printer owner has procured a corresponding blueprint from a 3D object designer. The 3D printer is controlled by personal computer that sends the G-code commands via a USB connection.

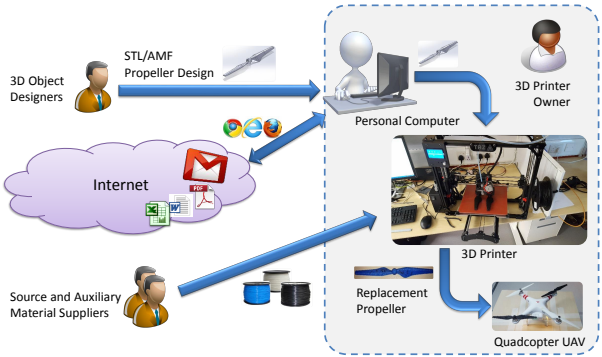

Figure 7. Considered Scenario

| DRONE | DJI Phantom 2 Vision+ |
|---|---|
| 3D PRINTER | LuLzbot Mini TAZ |
| FIRMWARE | Marlin 2015Q2 |
| CONTROLLER PC | Intel Xeon CPU, 32GB RAM, Windows 7 |
| SOFTWARE | ▪ SolidWorks<br>▪ Cura 19.12 |

TABLE 1. EXPERIMENTAL ENVIRONMENT, SUMMARY

This scenario's workflow is quite similar to the general AM workflow described earlier (see Figure 1 in Section 3), with just a few differences which will affect the selection of available attack vectors discussed later in this section. First, we assume that the considered 3D printer owner does not routinely maintains all of his software up-to-date, behavior which is representative for the majority of private desktop 3D printer owners, who either don't possesses the necessary skills, or are very reluctant to apply patches. Second, the printer owner also uses this PC to surf in the Internet, read e-mails, download documents, play games, etc.

In this scenario an adversary wants to cause a significant damage to the victim's drone. The adversary comes up with an idea to sabotage the victim's 3D-printed replacement propeller. This must be accomplished in such a way that during installation the propeller's integration in the drone is not compromised and that the change made to the replacement propeller is subtle enough to pass basic visual inspection. Furthermore, in the hope that the failure occurs while the drone reaches high altitude or during ascending at a high speed, the adversary needs to introduce a defect that will be time-delayed or triggered under certain operational conditions, respectively. We assume that the adversary has basic to average hacking skills but moderate to advanced AM proficiency. Additionally, the adversary is capable to procure the same desktop 3D printer and drone, in order to prepare and test attacks.

**6.1.2. Experimental Environment.** We have implemented the scenario outlined above using equipment that is somewhat typical for a private household (see summary in Table 1). The DJI Phantom 2 Vision+ is a 4 kg (8.8 pounds) quadcopter UAV that can stay in the air for up to 25 minutes. It has a HD video camera installed and can stream live video. The drone can be purchased for about $500, a price that is affordable for the majority of private users. For a 3D printer we have selected the LulzBot Taz 5 with single extruder and V2C connector. LuLzbot TAZ is a plug-and-play device that comes with a numerous of features that can make it attractive for a rather inexperienced user (e.g., auto-nozzle cleaning). LuLzbot implements Fused Deposition Modeling (FDM) technology which is characteristic for the bulk of the desktop 3D printers that are currently available. It can be purchased for about $1,250, placing it at the rather lower end of the desktop 3D printers. The 3D printer operates using *Marlin* firmware. Its counterpart, installed on the Controller PC, is *Cura*. Both are widely used and are recommended by several desktop 3D printer OEMs. The communication between the Controller PC and the 3D printer is established via a direct USB connection.

It should be noted that the LuLzbot 3D printer can operate using filaments of different materials as source materials. For printing propellers, we have selected Acrylonitrile Butadiene Styrene (ABS). It is a very durable, strong, flexible, shock absorbent, and heat resistant polymer – all properties important for our scenario. The latter is especially important to withstand the heat generated by the quadcopter's motors.

While being out of scope of this paper, it should be noted that this the manipulations or replacement of the source material represents a physical attack that can sabotage the manufactured part's quality [11], [13]. We have performed several laboratory experiments with propellers printed with Polylactic acid (PLA), another thermoplastic that is frequently used in 3D printing. During these experiments we have observed that PLA's relatively low melting point at about (approximately 150-160°C) does not withstand the heat that was generated by the Phantom's motors.

### 6.2. Attack Chain Selection

While our discussion in Sections 4 and 5 was of more general nature, in the current section we apply the developed concepts to identify an attack chain for the scenario presented (the selection is indicated in Figure 8).

**6.2.1. Attack Target.** In this scenario, the goal is for the propeller to break when the drone ascends very quickly or after a comparatively short operational time. This means that the sabotage of the propeller should either cause a reduction in the propeller's tensile strength in the direction of the lift force, or cause the rapid development of material fatigue in the propeller.

**6.2.2. Manipulations.** Out of the options described in Sections 4, the manipulation of the manufacturing process requires a higher degree of the AM proficiency. As for our scenario it is not the case, this option is eliminated. Both possible influences on object specification are feasible with the skills and tools our adversary has. This manipulation is preferable in the examined scenario since it is both selective and direct. However, a change to the orientation of the object can eventually have the following drawbacks. First of all, if the victim has already printed a propeller in the past (which is our assumption) and monitors the printing process, change of the printed object's orientation can easily be recognized. Furthermore, it is also possible that the object with changed orientation will not fit in the 3D printer platform, thus eventually preventing the propeller from being printed at all. The remaining option of defect introduction is both feasible and eventually can remain undetected.

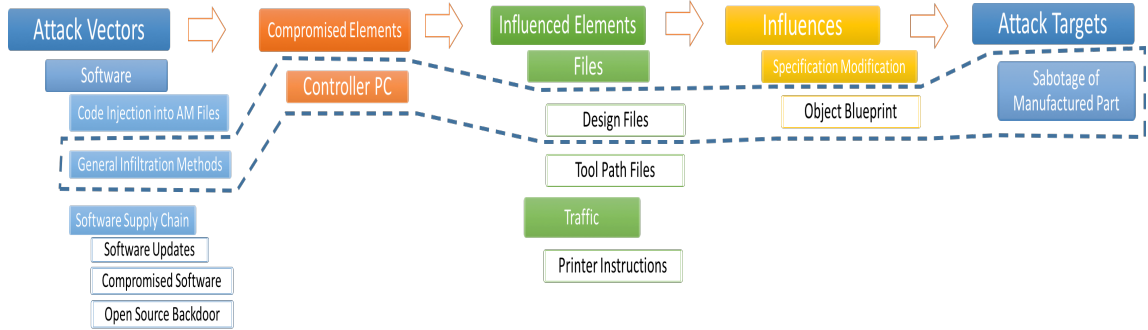

Figure 8. Case study attack chain

While the selection of the manipulation category is comparatively straight forward, the exact definition of the modification is not. Questions such as what type of defect, the shape of a defect, and its location in the printed 3D object should be defined.

Even though the adversary has moderate AM proficiency; it is extremely difficult to calculate impact of of defect's geometry on the stress concentration or similar – factors intensively studied and well understood in the material science. Therefore, the adversary has to experiment with various defects and find which are working within defined conditions.

Each defect can be introduced via multiple influenced elements. The choice of the influenced elements depends on the compromised element that is selected.

**6.2.3. Compromised Element.** The object description changes its representation as it traversal over different elements of workflow. It is represented as a STL/AMF file by the 3D object designer and during transfer to the Controller PC. Once at the Controller PC it is translated to either G-code individual commands or a toolpath file that are transmitted to the 3D printer. Also, in the 3D printer itself, the design can have a different internal representation. It is clear that any element along this path, if compromised, can modify the object specification.

In our considered scenario we aim to compromise the AM environment from the Internet, thus the Controller PC, the element that is both used for AM and surfing the Internet, is the valid choice.

**6.2.4. Attack Vectors.** Personal computers are a frequent targets of attacks. Numerous tools and relevant information can be found on the web. For choosing the attack vector, we have focused on most common tools and behaviors of individual users. We chose a phishing attack because it is one of the most common attack vectors for individuals and IT professionals. We had a choice of several file format or flash exploits that are widely available on the Internet. We chose a ZIP file format vulnerability in the popular WinRAR software.

### 6.3. Defining Manipulation

Regardless of adversary's AM proficiency, some imperial experiments are required to verify that the manipulation leads to the sabotage of the propeller. We assume that the adversary has enough resources to perform some experiments, using equipment either identical or close to identical to the victim's one. In particular, we assume that the adversary has the same 3D printer and drone. By gaining access to victim's Controller PC, the adversary possesses the original design file of the propeller and the knowledge of which slicing software is used. Figure 9 shows the original propeller design. It was modeled using the SolidWorks software. While playing the role of an adversary who is modifying the design, we use the SolidWorks software as well.

**6.3.1. Location of Defect.** Propellers convert engine's torque force into thrust force. The original propeller design is engineered to withstand the thrust force that acts on it when the motors are spinning at the maximum supported revolution per minute (RPM). The lift force causes the propeller to bent upwards during flight, thus placing a certain amount of stress onto the propeller.

The calculation of thrust force and stress distribution are rather complex topics[12] that, as we assume, exceed by far the knowledge of an average adversary. Nevertheless, even with a basic physics knowledge it is obvious that the biggest force is applied at the joint connecting the blades and the cap of the propeller together (see Figure 10). Therefore, this joint appears to be a good location to insert a defect in the design. Furthermore, if propeller breaks at this joint, the loss of thrust will be the maximal compared to all other places where a defect can be introduced.

---
12. The thrust produced by a propeller depends on factors like its shape and diameter, on the motor's RPM, and on the density of the air. Stress distribution depends on the object's geometry and material properties.

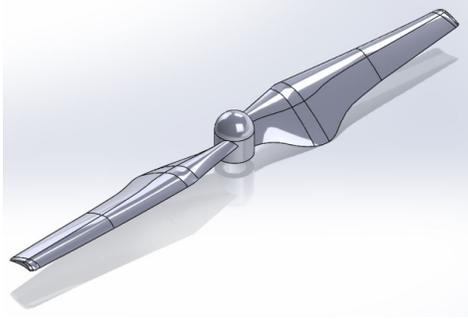

Figure 9. Original Propeller Designed using SolidWorks

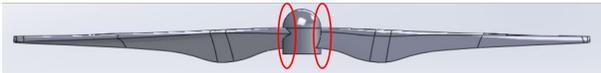

Figure 10. Red circles highlight location selected for the defect

**6.3.2. Iteration through Manipulations.** We have started investigation of possible manipulations by physically drilling holes between the blades and the cap after of a propeller printed using unaltered design. Perhaps a counter intuitive[13] outcome was that it has not really affected the propeller's ability to operate at the motor's maximum speed without breaking.

In the next phase, we have modified the design file by inserting internal rectangular gaps into the joint. We have iterated with different sized of the gap verified their impact empirically (see Figure 11). We found that 0.1mm length gaps to be optimal. Any gaps that were smaller did not influence the printing process enough to be meaningful. Larger gaps caused the propeller to break within seconds of normal activity. Moreover, we noticed that the introduction of gaps also influences the printing pattern of the joint between the cap and the blades. Figure 12 shows that the attachment surface of the blade to the cap is facing outwards in a normal design and inwards in a modified design. This modification to the attachment surface weakens its strength and accelerates fatigue. In order to reinforce the joint between the cap and the blades, we added internal support structure inside the gaps that connects the blades to the cap. During this iterative process, we have identified a modification that yielded satisfactory results, *i.e.,* the propeller broke after operating for a period of time.

---

13. While being counter intuitive, the outcome is based on the way how geometrical properties affect stress concentration/distribution. Geometrical discontinuities like cracks or sharp corners cause stress concentration; round cavities might, in opposite, contribute to the stress distribution.

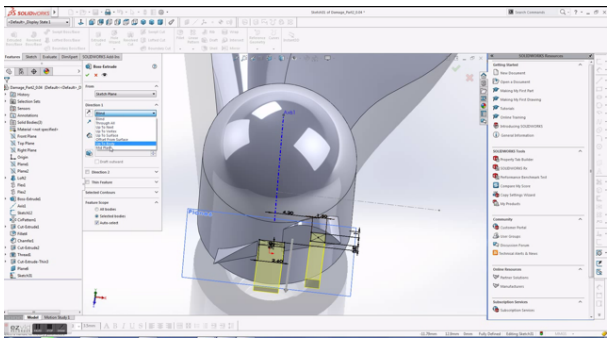

Figure 11. Propeller design with introduced defects

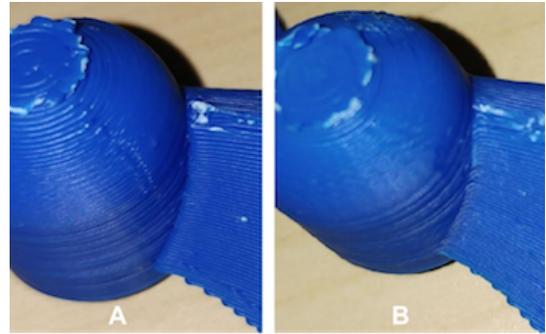

Figure 12. Two printed caps site-by-site. Cap A is *sabotaged* and Cap B is *benign*

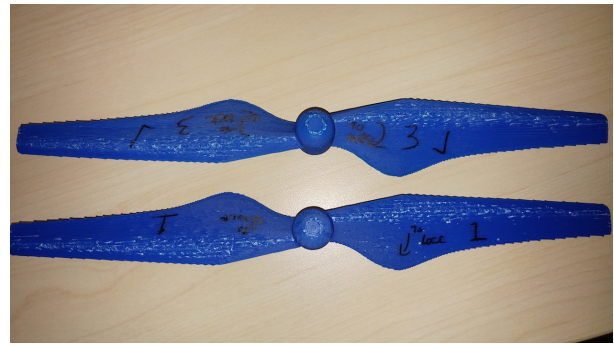

Figure 13. Two printed propellers site-by-site. The Upper is *benign* and the lower is *sabotaged*

It should be noted that, because the effects we have introduced are internal, the sabotage would remain unnoticed by a simple visual inspection of the printed propeller. Figure 13 shows two propellers side-by-side. The propeller at the top was printed using the original design, the propeller at the bottom using our maliciously altered design.

**6.3.3. Evaluation of Effects.** For every altered design, we have printed a propeller. We have evaluated the effects of introduced defects empirically, in laboratory settings. During all experiments we have always installed two 3D-printed propellers, one of the original and another one of a maliciously altered design.

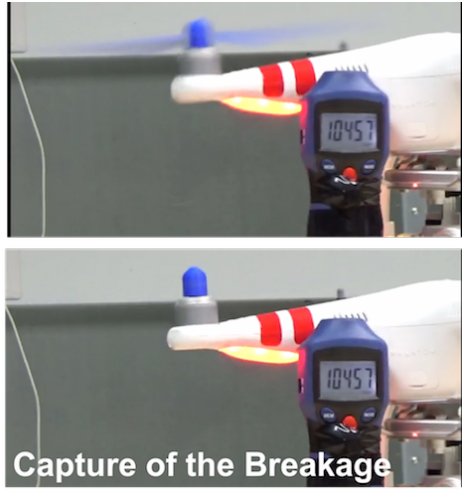

Figure 14. The damaged propeller just after it breaks in the lab

In order to create a controlled, reproducible testing environment and to measure effects of our manipulations, we have used an enclosed room and attached drone to the table with a duct tape. We have rapidly increased thrust/RPM to capture the breaking point. Doing this, we have measured two factors: the breakage time of the propeller and the RPM count when the failure happens.

During our tests we could verify that the propeller printed using the original design could operate at more than 15000 RPM for an extended period of time (over 5 minutes). However, propellers printed with various defects introduced in the design could not sustain this operational conditions. Figure 14 captures the second a sabotaged propeller breaks. The breakage point occurred at 10457 RPM after ∼10 seconds.

During our lab tests we have witnessed a side effect of the targeted damage that we aimed to create. Besides compromising the drone itself, in some cases the breakage of the propeller have damaged one of the other fully functioning propellers.

## 7. Case Study, Part II: Attack Execution

After all the above described preparatory steps have been accomplished, we were ready to perform the actual attack.

### 7.1. Sabotage Attack

We execute the sabotage attack in three steps. First, we have compromised the victim's Controller PC. Second, we have downloaded the original design file and developed a sabotaged design (as described in Section 6.3). Third, we have changed the design file on the victim's PC according to the developed manipulation.

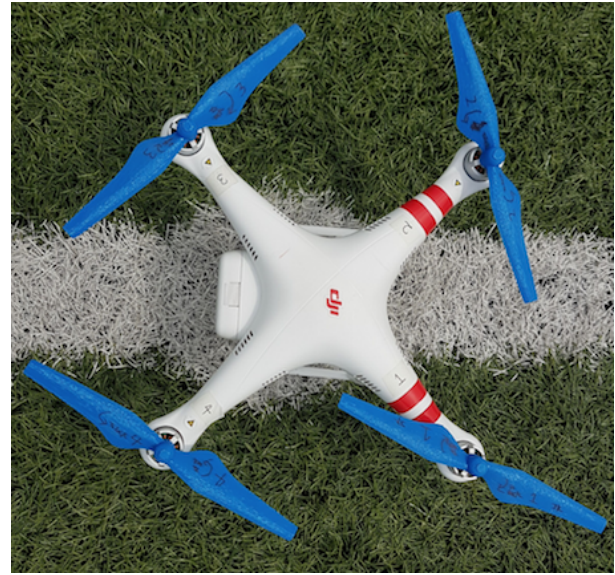

Figure 15. The drone with three normal propellers and one damaged

**7.1.1. Compromising Controller PC.** In order to infiltrate the system, we used a patched WinRAR vulnerability [35] that spoofs the file name and extension of the archived document. We have created a malicious EXE file using the Metasploit framework that triggers an exploit. Using the WinRAR vulnerability, we have changed the name and extension of the executable file to look like an innocent PDF file.

The victim received an email enticing him to download a .Zip file from Dropbox and double-click on the PDF file inside. Once the victim clicks on the file, a reverse shell is opened in the background and the attacker can take control over the system.

**7.1.2. Manipulating Design File.** The next step is to find potential design files that the attacker can manipulate. The attacker searches for .STL files and downloads them for further investigation. Once the files are in the attacker's possession he/she can modify the design in a way that would cause damage. In our experiment, we have modified the design using the SolidWork software, tested the effect of the manipulation experimentally (Section 6.3), and lastly, replaced the original file with the maliciously altered one. After the modified file is used to produce a replacement propeller, the sabotage attack enters a new phase, when the entire CPS can become a victim of a part failure.

### 7.2. Field Trials

The first step of the field trials was to test whether the printed propellers will be able to withstand actual flight. We attached four propeller that were all printed from the regular design. The drone was able to take off and fly normally during a long periods of time (more than 5 minutes of regular flight).

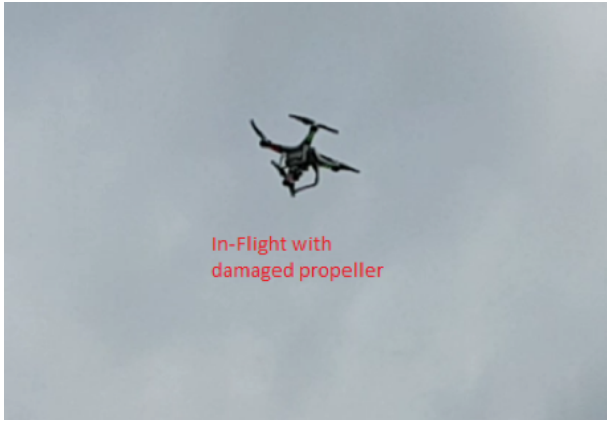

Figure 16. Drone in flight with the damaged propeller

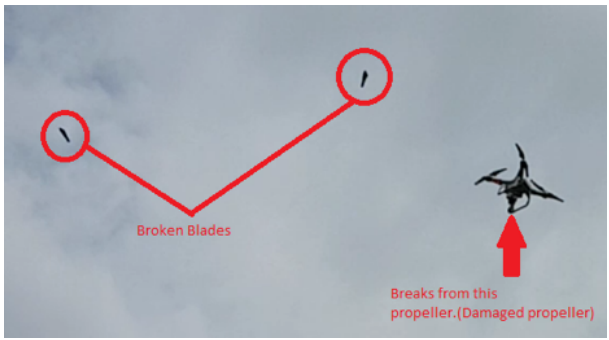

Figure 17. The damaged propeller breaks mid-flight

Then, to complete the attack, we have replaced a single propeller with a sabotaged 3D printed one. Figure 15 shows a drone with four 3D printed propellers attached. All propellers were printed from the original design file except one that was printed from the modified design file. During the final stress test, the propeller withstood a flight test of 1 min and 43 seconds (Figure 16). During this time, the drone was first ordered to accelerate from low to high altitude in short period of time. Then it has performed three cycles of low to high altitude changes. During these initial cycles, the sabotaged propeller performed normally. However, during the fourth iteration of rapid ascension the propeller broke apart.

When the propeller blades broke apart during the flight (see Figure 17), the drone was at the peak of its upward acceleration. Consequently, the quadcopter fell from a considerable height and shattered. The drone was severely damaged from that fall. The damage included one of the motors and the camera being completely destroyed, and the external casing of the drone cracked. We demonstrated how minor modification in the object's specification that were remotely introduced into the AM environment, caused the complete destruction of a cyber-physical system (see Figure 18).

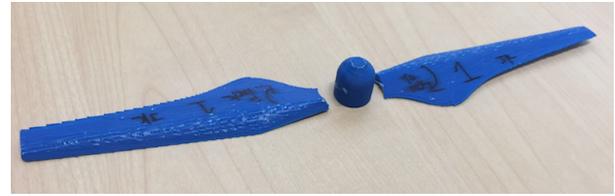

Figure 18. The broken sabotaged propeller

## 8. Conclusion

Additive Manufacturing (AM) emerges as a transformative manufacturing technology that is increasingly used for production of functional parts, including those for safety-critical systems. Due to the computerization of AM, several researchers have raised concern of its possible sabotage [9], [10], [11], [12], [13], [14]. Researchers have discussed attack vectors [12], [16], analyzed which AM manufacturing parameters can sabotage a part's quality [11], [29], shown experimentally that introduction of defects can degrade a part's mechanical properties [9], [14], and even speculated about possibility of weaponizing the AM process [13].

While being important contributions, prior works covers selected aspects of a possible attack only. To our best knowledge, this paper presents the very first full chain of attack involving AM, beginning with a cyber attack to compromise AM equipment, through a malicious modification of a design file, leading to the manufactured functional part's sabotage, and resulting in the physical destruction of a cyber-physical system that employs this part. For doing this, we have proposed a systematic approach to identify options for an attack involving AM that will allow an adversary to achieve his/her goals. Then we introduced a methodology for the assessment of the attack difficulty, thus enabling differentiation between possible attack chains. While this assessment is provided from the adversary perspective, it also allows a defender to reason about the most likely attacks. Finally, we applied the proposed approach on a selected scenario of a desktop 3D printer used to manufacture propellers of a quadcopter UAV. We show experimentally the proof for the whole attack chain. The video of the quadcopter's final flight when the sabotaged propeller broke can be found under [34].

Even though our experimental verification has compromised a private person's desktop 3D printer, we argue that similar attacks are possible on industrial systems producing metal parts for safety-critical systems. Therefore, in order to protect public safety and national security, solutions should be found and implemented that will increase both robustness and resilience of AM to sabotage attacks.